\documentclass[a4paper,dvipdfmx]{article}
\usepackage{ifthen}
\newcounter{IsDraftmode}
\ifthenelse{\value{IsDraftmode} < 1}{%
\usepackage{INTERSPEECH_v2}
}{}

\usepackage{amsmath}
\usepackage{amssymb,bm}
\usepackage{graphicx}
\usepackage{textcomp}
\usepackage{hyperref}
\usepackage[linesnumbered,lined,boxed]{algorithm2e}
\usepackage{color}
\usepackage{cite}

\ifthenelse{\value{IsDraftmode} < 1}{%
\sloppy 
\ninept

}{%

}

\title{A new cosine series antialiasing function and its application to \\
aliasing-free glottal source models for speech and singing synthesis\thanks{%
The main body of this article is accepted for publication in Interspeech2017.
This article has supplemental materials for details which were dropped due to limited space.}}
\ifthenelse{\value{IsDraftmode} < 1}{%
\name{Hideki Kawahara$^1$, Ken-Ichi Sakakibara$^2$, Masanori Morise$^3$, Hideki Banno$^4$, Tomoki Toda$^5$\\ Toshio Irino$^1$}
\address{
  $^1$Wakayama University, Japan\\
  $^2$Health Science University of Hokkaido, Japan\\
  $^3$Graduate School of Science and Technology, Meijo University, Japan\\
  $^4$Interdisciplinary Graduate School of Medicine and Engineering, University of Yamanashi, Japan\\
  $^5$Graduate School of Information Science, Nagoya University, Japan}
 \email{kawahara@sys.wakayama-u.ac.jp, kis@hoku-iryo-u.ac.jp, banno@meijo-u.ac.jp, mmorise@yamanashi.ac.jp, tomoki@icts.nagoya-u.ac.jp, irino@sys.wakayama-u.ac.jp}
}{%
\author{Hideki Kawahara, et.al.}
\date{\today}
}

\begin{document}

\maketitle
\begin{abstract}
We formulated and implemented a procedure to generate aliasing-free excitation source signals. 
It uses a new antialiasing filter in the continuous time domain followed by an IIR digital filter for response equalization.
We introduced a cosine-series-based general design procedure for the new antialiasing function.
We applied this new procedure to implement the antialiased Fujisaki--Ljungqvist model. 
We also applied it to revise our previous implementation of the antialiased Fant--Liljencrants model.
A combination of these signals and a lattice implementation of the time varying vocal tract model provides a reliable and flexible basis to test
$f_{\rm o}$ extractors and source aperiodicity analysis methods.
MATLAB implementations of these antialiased excitation source models are available as part of our open source tools for
speech science.
\end{abstract}

\section{Introduction}

Voice quality plays important roles in speech communication, especially in para- and non-linguistic aspects.
To test such aspects of speech communication, it is important to use relevant test stimuli that sound natural to listeners
and that at the same time have to be precisely determined and controlled. 
In addition, it is desirable for the stimuli to be easy to interpret in terms of voice production, as well as auditory perception.

Source filter models with source filter interaction\cite{titze2008Jasa} and glottal excitation models\cite{rosenberg1971effect,hedelin1984glottal,fant1985four,fujisaki1986icassp,fujisaki1987icassp,Klatt1990jasa,shue2010new,Alku2011}
 may provide practical and useful tools.
However, because glottal excitation comprises several types of discontinuity,
aliasing introduces spurious signals that interfere with reliable subjective tests.
This paper introduces a systematic procedure to eliminate the aliasing problem by
deriving a closed form representation of the antialiased excitation. 
We also introduce a new set of antialiasing functions using a cosine series to keep
the level of spurious signals around the Nyquist frequency low.
We will make them accessible by
providing MATLAB implementations as well as interactive GUI-based tools\cite{kawahara2016isST}.

\section{Background and related work}
One of the authors has been developing a speech analysis, modification and resynthesis framework and related tools\cite{kawahara1999spcom,kawahara2008icassp,kawahara2013apsipa}.
They are based on $f_{\rm o}$\footnote{We use ``$f_{\rm o}$'' instead of ``F0'' based on Reference\cite{titze2015jasaforum}.
}-adaptive procedures, which require reliable and precise extraction of $f_{\rm o}$ and aperiodicity information.
Development of such $f_{\rm o}$ extractors requires dependable ground truth.
The aliasing-free L--F model (Fant--Liljencrants model) provided the ground truth for developing our new source information analysis framework\cite{kawahara2016using}.

A search for aliasing-free glottal source models produced one reference\cite{milenkovic1993voice}, which is not directly applicable to antialiasing of the L--F model
for two reasons.
First, it provides a procedure to antialias a piece-wise polynomial function, whereas the L--F model is a piece-wise exponential function.
We had to derive the closed-form representation of the antialiased L--F model by ourselves\cite{kawahara2015apsipa}.
Second, the reference by Milenkovic\cite{milenkovic1993voice} had some typos, some equations were missing and a sample implementation was not available.
In this article, we repaired the procedure and developed executable MATLAB functions.
Retrospectively we found BLIT (band-limited impulse train)-based methods\cite{stilson1996alias,valimaki2007ieeesm}.
We found that
literatures on digital representation of analog musical signals also provide aliasing reduction methods\cite{valimaki2005dtms,Valimaki2005ieee,pekonen2011brief}.
Our formulation 
with 
a new cosine series
is more flexible and provides better aliasing suppression.

There are two important reasons for deriving an aliasing-free Fujisaki--Ljungqvist model.
First reason is the relevance of the model.
We found that the L--F model does not necessarily model the actual glottal source behavior, especially extreme voices\cite{sakakibara2011physiological}.
Such voices sometimes consist of stronger discontinuities than the L--F model provides.
The Fujisaki--Ljungqvist model provides several different levels of discontinuity and is reported to fit actual speech samples better\cite{fujisaki1987icassp}.

The second and the most important reason is that it enables to develop a general procedure for antialiasing glottal source models.
The Fujisaki--Ljungqvist model is a piece-wise polynomial,
whereas the L--F model is a piece-wise exponential.
The other popular glottal source models can be represented using both or one of these representations\cite{rosenberg1971effect,hedelin1984glottal,fant1985four,fujisaki1986icassp,fujisaki1987icassp,Klatt1990jasa,shue2010new,Alku2011}.
Developing a procedure to make the Fujisaki--Ljungqvist model aliasing-free provides the necessary means of attaining this goal.


\section{Fujisaki--Ljungqvist model}
The excitation signal $E(t)$ of the Fujisaki--Ljungqvist model  is defined by the following equation in the continuous time domain\cite{fujisaki1986icassp,fujisaki1987icassp}:
\vspace{-0.2cm}
{\small
\begin{align}
\label{ eq:flmodel}
\! \! \! \!E(t) & = \left\{\begin{array}{ll}
\displaystyle 
\! \! \!  A - \frac{2A + R \alpha}{R}t + \frac{A + R\alpha}{R^2}t^2  ,     & \! \! \! \! \! \! \! \!  0 < t \le R \\ [7pt]
\displaystyle \! \! \!  \alpha(t-R) + \frac{3B-2F\alpha}{F^2}(t-R)^2 & \\ [5pt]
\displaystyle {} \ \ \ \ \ \ -\frac{2B-F\alpha}{F^3}(t-R)^3  ,   &  \! \! \! \! \! \! \! \!   R < t \le W \\ [6pt]
\displaystyle \! \! \! C - \frac{2(C-\beta)}{D}(t-W) & \\ [5pt]
\displaystyle {} \ \ \ \ \ \ + \frac{C-\beta}{D^2}(t-W)^2 , & \! \! \! \! \! \! \! \! \!  \!  \! W\!<\!t \!\le\! W\!+\!D \\ [6pt]
\! \! \! \beta  , & \! \! \! \! \! \! \!  \!\! \! \!   W\!+\!D\! <\! t\! \le \!T
\end{array}
\right. \! \! \! \! \! \! \! \! \! ,
\end{align}
}
\vspace{-0.4cm}
where
{\small
\begin{align}
\alpha & = \frac{4AR-6FB}{F^2-2R^2} \ \ \mbox{and} \ \ \ \beta = \frac{CD}{D-3(T-W)} .
\end{align}
}
\vspace{-0.2cm}

\begin{figure}[tbp]
\begin{center}
\includegraphics[width=0.99\hsize]{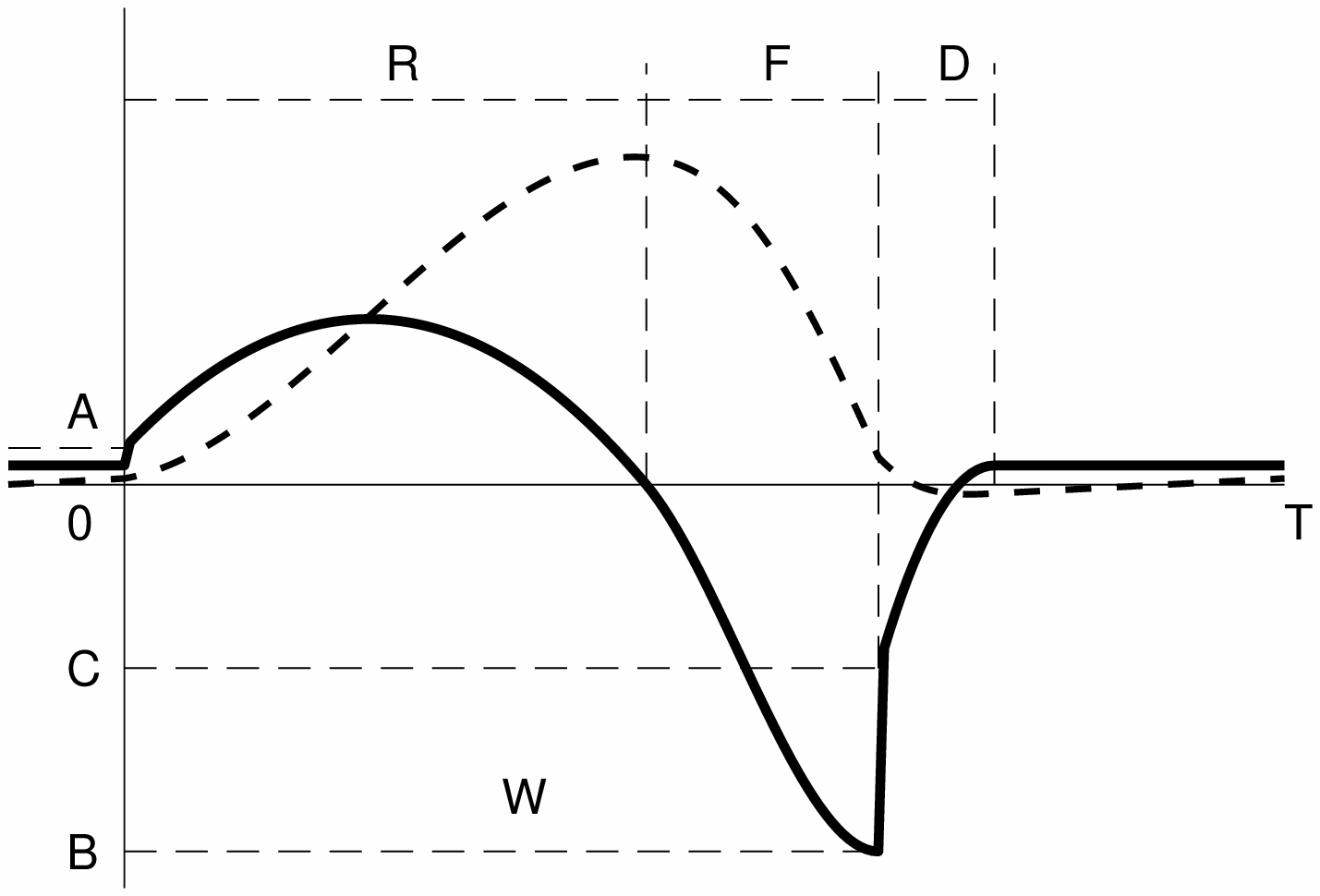}
\vspace{-0.3cm}
\caption{Fujisaki--Ljungqvist model.
Glottal air flow $U_g(t)$ (dashed line) and the excitation signal
$E(t)$ (solid line)}
\label{flModelShape}
\end{center}
\vspace{-0.3cm}
\end{figure}
Figure~\ref{flModelShape} shows the Fujisaki--Ljungqvist model parameters and waveforms.
The excitation signal $E(t)$ is a time derivative of the glottal flow $U_g(t)$.
When the fundamental period $T$ is normalized to one,
six model parameters determine the shape.
They are $A, B, C, R, F,$  and $D$.

The Fujisaki--Ljungqvist model is a piece-wise polynomial function.
Because filtering for antialiasing is a linear operation,
antialiasing of the Fujisaki--Ljungqvist model is solved by adding each
antialiased polynomial pulse $p(\tau)$, which is defined in $(0 \le \tau \le 1)$ using relevant scaling:
\vspace{-0.2cm}
\begin{align}
p(\tau) & = p_0 \tau^0 + p_1 \tau^1 + \cdots + p_n \tau^n , \ \ 0 \le \tau \le 1 
\end{align}
where the scaling is given by $\tau = t/T$.

\section{Antialiasing a polynomial pulse}
We use the framework proposed by reference\cite{milenkovic1993voice}\footnote{%
Note that we removed the model weighting coefficient $w_m$ in the original reference\cite{milenkovic1993voice}.
It significantly simplifies the following discussion.
}.
It starts from a matrix representation by introducing step functions and antialiasing in the continuous time domain,
followed by equalization in the discrete time domain.
The simplification and this equalization, as well as their application to the Fujisaki--Ljungqvist model, are our contribution.

This section fixes typos, adds missing equations and clarifies and adds missing conditions in the original reference\cite{milenkovic1993voice}.
All equations are defined in the continuous time domain.

\subsection{Matrix form}
Let us start with the introduction of the two step functions $u(\tau)$ and $u(\tau-1)$,
which yields
\vspace{-0.2cm}
\begin{align}
p(\tau) & = p(\tau) u(\tau) - p(\tau) u(\tau -1)  \nonumber \\
& = \mathbf{p}^T \mathbf{u}_{\tau} - \mathbf{p}^T \mathbf{B} \mathbf{u}_{\tau - 1} ,
\end{align}
where $\mathbf{B}$ is a lower triangular matrix defined as follows:
\begin{align}
\mathbf{B} = & \left[
\begin{array}{llll}
 b_{00}     &    0 & \cdots & 0 \\ [-4pt]
 b_{10}     &  b_{11}  & \ddots & \vdots \\ [-4pt]
 \vdots & \ddots  & \ddots & 0 \\
 b_{n0} & b_{n1} & \cdots & b_{nn}
\end{array}
\right] ,
\end{align}
where the coefficient $b_{nk} = n! / (k! (n-k)!)$ is a binomial coefficient,
and $\mathbf{p}$,  $\mathbf{u}_{\tau}$, and  $\mathbf{u}_{\tau-1}$ are vectors defined as follows:
\vspace{-0.2cm}
{\small
\begin{align}
\!\!\!\mathbf{p} = & \!\left[\!\!\!
\begin{array}{l}
    p_0    \\
     p_1 \\
      \vdots \\
     p_n
\end{array}
\!\!\!\! \right]\!\! , 
\mathbf{u}_{\tau} =\!
\left[\!\!\!
\begin{array}{l}
      \tau^0 u(\tau)    \\
       \tau^1 u(\tau) \\
      \vdots \\
       \tau^n u(\tau)
\end{array}
\!\!\!\!\right]\!\! , \ \
\!\!\mathbf{u}_{\tau-1} =\!\!
\left[\!\!\!
\begin{array}{l}
      (\tau-1)^0 u(\tau-1)    \\
       (\tau-1)^1 u(\tau-1) \\
      \vdots \\
       (\tau-1)^n u(\tau-1)
\end{array}
\!\!\!\!\right] 
\end{align}
}

\subsection{Antialiasing using a cosine series}
The antialiased filtered pulse $p_h(\tau)$ has the following form:
\begin{align}\label{eq:tobeDiscre}
p_h(\tau) & = h(\tau) \ast \{ p(\tau) u(\tau) - p(\tau) u(\tau -1) \} ,
\end{align}
where $h \ast u$ represents the convolution of $h$ and $u$.
Convolution is a linear operation and yields the following:
\begin{align}
p_h(\tau) & = \mathbf{p}^T \mathbf{h}_{\tau} - \mathbf{p}^T B \mathbf{h}_{\tau - 1} , \label{eq:aapolyFirst}
\end{align}
where $\mathbf{h}_{\tau}$ and $\mathbf{h}_{\tau - 1}$ represent  the following vectors:
{\small
\begin{align}
\!\!\!\!\!\mathbf{h}_{\tau}\!  = \!\!
\left[\!\!\!\!
\begin{array}{l}
      h(\tau) \!\ast\! \tau^0 u(\tau)    \\
      h(\tau) \!\ast \!\tau^1 u(\tau) \\
      \vdots \\
      h(\tau) \ast \tau^n u(\tau)
\end{array}
\!\!\!\!\right]\!\! , 
\mathbf{h}_{\tau-1}\! = \!\!
\left[\!\!\!
\begin{array}{l}
      h(\tau) \!\ast\! (\tau\!-\!1)^0 u(\tau\!-\!1)    \\
      h(\tau) \!\ast\! (\tau\!-\!1)^1 u(\tau\!-\!1) \\
      \vdots \\
      h(\tau) \!\ast\! (\tau\!-\!1)^n u(\tau\!-\!1)
\end{array}
\!\!\!\right]\!\! .
\end{align}
}

Equation~\ref{eq:aapolyFirst} provides the antialiased signal to be discretized. 
We use the following cosine series for antialiasing the filter response $h(t)$:
\begin{align}\label{eq:cosSerGen}
h(t) & = \sum_{k=0}^{m} h_k \cos\left(\frac{k \pi t}{t_w}\right) , & -t_w < t \le t_w ,
\end{align}
where $m$ represents the exponent of the highest order term and
$t_w$ represents the length of the filter, which is determined according to the sampling frequency.

Assigning specific values to the coefficients $h_k$ determines $\mathbf{h}_{\tau}$ and $\mathbf{h}_{\tau - 1}$.
We derive the closed-form representation of Equation~\ref{eq:tobeDiscre} and then discretize it.
Equations~\ref{eq:htv} and \ref{eq:htm1v} determine  $\mathbf{h}_{\tau}$ and $\mathbf{h}_{\tau - 1}$, respectively.
The next step is to determine the matrices $\mathbf{C}, \mathbf{S}, \mathbf{U}$, and $\mathbf{V}$, with constant coefficients.
\begin{align}
& \mathbf{h}_{\tau}  =
\left\{
\begin{array}{ ll}
 0   ,  &  t \le -t_w \\
 \mathbf{C} \mathbf{c}_t + \mathbf{S} \mathbf{s}_t + \mathbf{U} \mathbf{t}_{n+1},   &   -t_w < t \le t_w \\
 \mathbf{V} \mathbf{t}_n  , &   t_w < t
\end{array}
\right.  \label{eq:htv} \\
& \mathbf{h}_{\tau-1} \! =
 \mathbf{C} \mathbf{c}_{t-1} \!+\! \mathbf{S} \mathbf{s}_{t-1} \!+\! \mathbf{U} \mathbf{d}_{n+1}, \ \ 1\!-\!t_w \!<\! t \!\le\! 1\!+\!t_w ,
 \label{eq:htm1v} 
\end{align}
where each element of the vectors $ \mathbf{c}_t$, $ \mathbf{s}_t$, $ \mathbf{t}_{n+1}$, and $ \mathbf{t}_n$ is defined as follows,
where $(\mathbf{c}_t)_k$ represents the $k$-th element of $ \mathbf{c}_t$:
\begin{align}
\!\!\!(\mathbf{c}_t)_k = \cos\!\left(\!\!\frac{k \pi t}{t_w}\!\!\right) , \ (\mathbf{s}_t)_k = \sin\!\left(\!\!\frac{k \pi t}{t_w}\!\!\right) , \ (\mathbf{t}_n)_k = t^{k - 1}  ,
\end{align}
and vectors $ \mathbf{c}_{t-1}$, $ \mathbf{s}_{t-1}$, and $ \mathbf{d}_{n+1}$ are defined thus:
\begin{align}
(\mathbf{c}_{t-1})_k & = \cos\!\left(\!\!\frac{k \pi (t-1)}{t_w}\!\!\right) , \ (\mathbf{s}_{t-1})_k = \sin\!\left(\!\!\frac{k \pi {(t-1)}}{t_w}\!\!\right) \nonumber \\
(\mathbf{d}_{n+1})_k & = (t-1)^{k - 1}  .
\end{align}

\subsection{Recursive determination of coefficients}
The recursion uses the following continuity constraints and initial values. The continuity gives the following:
\begin{align}
&  \mathbf{C} \mathbf{c}_t + \mathbf{S} \mathbf{s}_t + \mathbf{U} \mathbf{t}_{n+1} = 0 & t = -t_w\\
&  \mathbf{C} \mathbf{c}_t + \mathbf{S} \mathbf{s}_t + \mathbf{U} \mathbf{t}_{n+1} = \mathbf{V} \mathbf{t}_n & t = t_w .
\end{align}

The initial condition is for the first row of the matrices, where $m$ represents the 
highest exponent of the selected cosine series antialiasing function.
In the following equation, $\mathbf{C}_{r,k}$ represents the element of the $r$-th row and $k$-th column:
\begin{align}
\begin{array}{l}
\mathbf{C}_{0,k} = 0,  \ \ 1 \le k \le m , \\ [4pt]
\displaystyle \mathbf{S}_{0,k} = \left(\frac{t_w}{k \pi} \right) h_k,  \ \ 1 \le k \le m , \\ [4pt]
\mathbf{U}_{0,1} = h_0, \\ [4pt]
\mathbf{U}_{0, k} = 0 , \ \ 1< k \le n+1 ,\\  [4pt]
\mathbf{V}_{0, 0} = 1, \ \ \mbox{(note that this is missing in \cite{milenkovic1993voice})} ,  \\ [4pt]
\mathbf{V}_{0, k} = 0, \ \ 1 \le k \le n ,
\end{array} \label{eq:defU01}
\end{align}

For $r=1$ through $r=n$ the following recursion determines the coefficients.
{\small
\begin{align}
\!\!\!\!\!\!\!\begin{array}{l}
\displaystyle \mathbf{C}_{r,k} = -\left(\frac{r t_w}{k \pi} \right) \mathbf{S}_{r-1, k} ,  \ \ 1 \le k \le m , \\  [8pt]
\displaystyle \mathbf{S}_{r,k} = \left(\frac{r t_w}{k \pi} \right) \mathbf{C}_{r-1, k} ,  \ \ 1 \le k \le m , \\ [8pt]
\displaystyle \mathbf{U}_{r, k} = \left(\frac{r}{k}\right)\mathbf{U}_{r-1, k-1} , \ \ 1\le k \le n+1 ,\\ [5pt]
\displaystyle \mathbf{U}_{r,0} = - \sum_{k=1}^{n+1} (-t_w)^{k} \mathbf{U}_{r,k} - \sum_{k=1}^m (-1)^{k} \mathbf{C}_{r,k},  \\ [8pt]
\displaystyle \mathbf{V}_{r, k} = \left(\frac{r}{k}\right)\mathbf{V}_{r-1, k-1} , \ \ 1\le k \le n , \\ [6pt]
\displaystyle \mathbf{V}_{r,0} = \!\!\sum_{k=0}^{n+1} (t_w)^k \mathbf{U}_{r,k} \! + \!\!\sum_{k=1}^m (-1)^k \mathbf{C}_{r, k} \!- \!\!\sum_{k=1}^{n} (t_w)^k \mathbf{V}_{r, k}, 
\end{array}
\end{align}}
where the fourth and sixth lines were placed in a confusing manner in Reference\cite{milenkovic1993voice}.
Further, the reference has a typo in Eq.~\ref{eq:defU01}, which defines $\mathbf{U}_{0,1}$.

\subsection{Closed form representation}
The following equation provides the antialiased polynomial value at a given $t$:
\vspace{-0.2cm}
\begin{align}\label{eq:aaPolyFinalFix}
\!\!\!\!p_h(t) & \!= \!\!\left\{\!\!
\begin{array}{ll }
\!\!\mathbf{c}_0 \mathbf{c}_t +    \mathbf{s}_0 \mathbf{s}_t  + \mathbf{u}_0 \mathbf{t}_{n+1} & \!\!\mathcal{Q}_1(t)  \\ [4pt]
\!\!\mathbf{v} \mathbf{t}_n      &  \!\! \mathcal{Q}_2(t)  \\ [3pt]
\!\!\mathbf{c}_0 \mathbf{c}_t +    \mathbf{s}_0 \mathbf{s}_t  + \mathbf{u}_0 \mathbf{t}_{n+1}  & \\
 \ \ \ - (\mathbf{c}_1 \mathbf{c}_{t-1} +    \mathbf{s}_1 \mathbf{s}_{t-1}  + \mathbf{u}_1 \mathbf{d}_{n+1})  &  \!\!\mathcal{Q}_3(t)    \\ [4pt]
\!\!\mathbf{v} \mathbf{t}_n  - (\mathbf{c}_1 \mathbf{c}_{t-1} +    \mathbf{s}_1 \mathbf{s}_{t-1}  + \mathbf{u}_1 \mathbf{d}_{n+1})  & \!\!\mathcal{Q}_4(t) 
\end{array}
\right. \\
& \mbox{where} \nonumber \\
& \begin{array}{l}
\mathcal{Q}_1(t) = \{  -t_w < t \le t_w \wedge t \le 1 - t_w \} \\
\mathcal{Q}_2(t) = \{  t_w < t \le 1 - t_w \} \\
\mathcal{Q}_3(t) = \{  1 - t_w < t \le  t_w \} \\
\mathcal{Q}_4(t) = \{  1-t_w < t \le 1 + t_w \wedge t_w < t \}
\end{array} ,
\end{align}
where $\wedge$ represents logical AND. 
Note that we have to refine the conditions given in Reference\cite{milenkovic1993voice} to make this procedure to work properly.
The coefficient vectors are defined as follows:
\vspace{-0.2cm}
\begin{align}\label{eq:spcoeff}
\!\!\!\!\!\!\!\!\!\begin{array}{llll}
\mathbf{c}_0\! =  \mathbf{p}^T \mathbf{C} , &\!\! \mathbf{s}_0\! =  \mathbf{p}^T \mathbf{S}, & \!\!\mathbf{u}_0\! =  \mathbf{p}^T \mathbf{U} , & \!\!\! \mathbf{v}\! = \mathbf{p}^T \mathbf{V} , \\
\mathbf{c}_1\! =  \mathbf{p}^T \mathbf{B} \mathbf{C} , & \!\! \mathbf{s}_1\! =  \mathbf{p}^T \mathbf{B} \mathbf{S} , & \!\! \mathbf{u}_1\! =  \mathbf{p}^T \mathbf{B} \mathbf{U} & 
\end{array}
\end{align}

\section{Antialiasing function}
We introduce two new cosine series antialiasing functions here.
Because of the infinite frequency range of the glottal excitation models, which have discontinuities,
commonly used time windowing functions with 6~dB/oct sidelobe decay\cite{slepian1961prolate,slepian1978prolate,harris1978ieee,kaiser1980use} introduce
significant spurious due to aliasing.
In our previous derivation, we used 
one of Nuttall's windows\cite{nattall1981ieee} for the antialiasing function.
%
The maximum side lobe level of the window is $-82.60$~dB and the decay speed of side lobes is 30~dB/oct.
This decay rate is not steep enough to suppress spurious components around $f_{\rm o}$,
and the sidelobe level is not low enough to suppress spurious components around the Nyquist frequency.
The other windows listed in \cite{nattall1981ieee} cannot solve both issues at the same time.

\subsection{Design procedure}\label{ss:design}
Using s similar process as that used in \cite{nattall1981ieee},
we design a new set of windows to satisfy both the sidelobe level and the decay conditions.
We use a cosine series to design the antialiasing function.
The elements of the cosine series have the following form:
\vspace{-0.2cm}
\begin{align}
\psi_k(t) & = \left\{
\begin{array}{ll}
\displaystyle \cos\left( \frac{k \pi t}{t_w} \right) ,  &   -t_w \le t \le t_w   \\ [4pt]
0  &  |t| > t_w 
\end{array}
\right. .
\end{align}

For the designed function to behave properly for antialiasing,
the following conditions have to be satisfied.
Note that the odd ordered derivatives are always zero for
$t = \pm t_w$.
\begin{description}
\item[Sum of coefficients should be one] 
This determines the height of the function at the origin.
For simplicity, we set it to one.
\item[Level at the end point should be zero]
The function should be continuous at the end point.
\item[Derivatives at the end point should be zero]
Depending on the required slope, the derivatives at the end point have to be equal to zero.
For the decay rate $6 + (12 \times P)$~dB/oct, derivatives up to the order $2P$ are zero.
\end{description}
These conditions are summarized by the following 
equations.
{\small
\begin{align}
\!\!h(0) & = \sum_{k=0}^{m} h_k = 1 \\
\!\!h(\pm t_w) & = \sum_{k=0}^{m} (-1)^{k} k^0 h_k = 0 \\
\!\!\left. \frac{d^{2p} h(t)}{d t^{2p}}\right|_{t = \pm t_w}\!\!\! & = \sum_{k=0}^{m} (-1)^{k} k^{2p} h_k= 0, \ p = 1, \ldots , P
\end{align}}

Once the desired decay is decided, it provides $P + 2$ conditions.
If the number of coefficients of the cosine series is equal to $P+2$, there is no room for adjustment.
By adding one adjustable coefficient $q_0$, we can control the sidelobe level.
It yields the following equation:
\vspace{-0.2cm}
\begin{align}
\mathbf{q} = \mathrm{R} \mathbf{g} ,
\end{align}
where
\vspace{-0.2cm}
\begin{align}
\!\!\!\mathrm{R}\!\! & = \!\!\!\left[\!\!
\begin{array}{cccccc}
1  &  1 & \cdots  &  1  & 1 & 1 \\
1  &  -1 &  \cdots &  (-1)^k k^0  & \cdots & (-1)^m m^0 \\
0  &  -1 & \cdots  & (-1)^k k^2   & \cdots & (-1)^m m^2  \\ [-4pt]
\vdots  & \vdots  & \vdots  &  \vdots  & \vdots & \vdots \\ 
0  &  -1 & \cdots  &  (-1)^k k^{2p}  & \cdots & (-1)^m m^{2p} \\  [-4pt]
\vdots  & \vdots  & \vdots  & \vdots   & \vdots & \vdots \\ 
0  & -1  &  \cdots  &  (-1)^k k^{2P}  & \cdots & (-1)^m m^{2P}\\
1  & 0  &  \cdots  &  0  &  \cdots & 0
\end{array}
\!\!\!\right] \\
\mathbf{q} & = [1, 0, \cdots, 0, q_0]^T .
\end{align}

The solution $\mathbf{g}$ is given by
\vspace{-0.2cm}
\begin{align}
\mathbf{g} & = \mathrm{R}^{-1} \mathbf{q} ,
\end{align}
where the elements of $\mathbf{g}$ provide the coefficients of the cosine series.
Designing an antialiasing function means tuning the parameter $q_0$ to minimize the target cost.
This time the target is the level of the maximum sidelobe level. 
For a 42~dB/oct decay, a five-term cosine series is designed.
For a 54~dB/oct decay, a six-term cosine series is designed.

Numerical optimization yielded the following coefficients:\footnote{%
The coefficients are rounded to ten digits to the right of the decimal point. The best coefficients were selected from rounded numbers.}
\vspace{-0.2cm}
{\small
\begin{align}
\!\!\!\{h_k\}_{k=0}^4  & =  
 \left\{0.2940462892, 0.4539870314, 0.2022629686, \right. \nonumber \\ 
&\!\!\!\!\!\! \left. {} 0.0460129686, 0.0036907422 \right\} \\
\!\!\!\{h_k\}_{k=0}^5 & =  \left\{0.2624710164, 0.4265335164, 0.2250165621,  \right. \nonumber \\ 
& \!\!\!\!\!\! \left. 0.0726831633, 0.0125124215, 0.0007833203\right\} , \label{eq:sixTerm}
\end{align}}
Substituting these coefficients in Eq.~\ref{eq:spcoeff} provides an antialiased Fujisaki--Ljungqvist model. 
Half of the window length $t_w$ is set to make the first zero of the frequency domain
representation coincide with half of the sampling frequency $f_s$. 

\begin{figure}[tbp]
\begin{center}
\includegraphics[width=0.99\hsize]{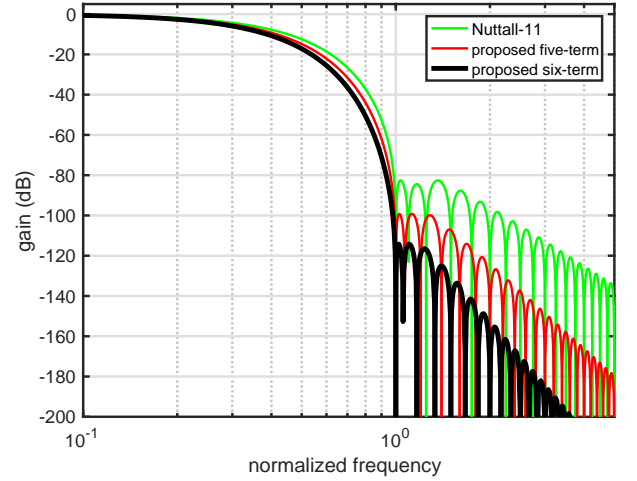}
\vspace{-0.3cm}
\caption{Gain of new antialiasing functions and the Nuttall-11 window.}
\label{newAAfunctions}
\end{center}
\vspace{-0.5cm}
\end{figure}
Figure~\ref{newAAfunctions} shows the gain of these new functions and
the Nuttall window that was used for the aliasing-free L--F model\cite{kawahara2015apsipa}.
The five-term function has  a $-99.23$~dB maximum sidelobe level with a 42~dB/oct decay.
The six-term function has a $-114.24$~dB maximum sidelobe level with a 54~dB/oct decay.
We decided to use the six-term function afterward.

\subsection{Discrete time domain: equalization}
The final stage, which is processed in the discrete time domain,
is equalization.

The designed antialiasing functions introduce severe attenuation around the Nyquist frequency.
In our antialiased L--F model, an FIR equalizer was designed to compensate for this attenuation\cite{kawahara2015apsipa}.
We used a
simple IIR filter with six poles in this Fujisaki--Ljungqvist model and found that it equalizes this attenuation effectively.
A sample implementation produces equalized gain deviations
from the FIR version within $\pm 0.2$~dB from 0 to 16~kHz for 44100~Hz sampling.

\section{Application to glottal source models}
The antialiased Fujisaki--Ljungqvist model output is obtained by the calculation of
each antialiased polynomial and sum together.
Discretization is the calculation of the antialiased Fujisaki--Ljungqvist model value at each sampling instance.
Finally,
applying the discrete time IIR equalizer to the discretized samples provides the discrete signal of the antialiased Fujisaki--Ljungqvist model.

\subsection{Examples: Fujisaki--Ljungqvist model performance}
\begin{figure}[tbp]
\begin{center}
\includegraphics[width=0.99\hsize]{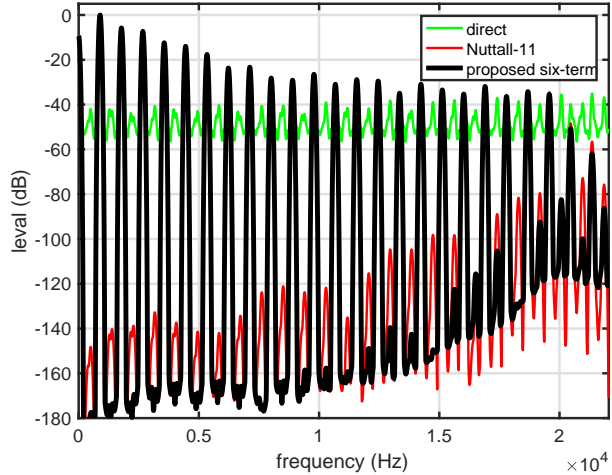}
\vspace{-0.3cm}
\caption{Power spectra of direct and two antialiased signals of
the Fujisaki-Ljungqvist model.
$f_{\rm o}$ is set to 887~Hz to make spurious components due to aliasing 
appear between harmonic components and look salient.
}
\label{aaFLspecSlice880Nuttall11}
\end{center}
\vspace{-0.5cm}
\end{figure}
We implemented this procedure using MATLAB and prepared high-level APIs.
One function generates an excitation source signal using the given $f_{\rm o}$ trajectory and
the time-varying Fujisaki--Ljungqvist model parameter set $A(t), B(t), C(t), R(t), F(t)$ and $D(t)$.

Figure~\ref{aaFLspecSlice880Nuttall11} shows the power spectra of the Fujisaki--Ljungqvist model outputs.
Direct discretization generates aliasing noise around $-60$~dB from the peak harmonics level.
The noise level of the final equalization around the fundamental component is about $-180$~dB
when using the six-term proposed function.
Antialiasing using the Nuttall-11 window introduces approximately 20~dB higher spurious levels.
(Note that the noise level using \texttt{nuttallwin} of MATLAB as antialiasing is approximately $-120$~dB. )

\subsection{Revision of antialiased L--F model}
We revised our previous antialiased L--F model\cite{kawahara2015apsipa} using the new six-term cosine series and
the six-pole IIR equalizer.
We also reformulated the algorithm using the element function $\varphi_k(t) = I_1+I_2+I_3$ as a building block
to define the antialiased and normalized complex exponential pulse $p_h(t)$:
\vspace{-0.2cm}
{\small
\begin{align}
p_h(t) & = \frac{1}{2 t_w h_0}\sum_{k=0}^{6} h_k \varphi_k(t) ,
\end{align}}
where $\{h_k\}_{k=0}^{6}$ is given by Eq.~\ref{eq:sixTerm}.
The factor $2 t_w h_0$ is for the gain normalization.
For each $k>0$, the explicit form is as follows:
\vspace{-0.2cm}
{\small
\begin{align}
\!\!\! I_1 & = \frac{k\alpha\sin(k\alpha t) \! - \! \beta\cos(k\alpha t) \!+\!(\!-1\!)^k \beta\exp(\beta( t_w\!\!+\!t))}{k^2\alpha^2 + \beta^2} \\
\!\!\! I_2 & = \frac{(-1)^k\beta\exp(\beta t) (\exp(\beta t_w)-\exp(-\beta t_w))}{ k^2\alpha^2 + \beta^2} \\
\!\!\! I_3 & = \frac{1}{ k^2\alpha^2+\beta^2} \{ k\alpha\exp(\beta)\sin(k\alpha (t - 1))   \nonumber \\
&   {}   -\! \beta\exp(\beta)\cos(k\alpha (t \!- \!1))\!  +\! (\!-1\!)^k\beta\exp(\beta (t_w\!+ \! t)) \} ,
\end{align}
where $\alpha=\pi/t_w$. 
The functions $I_1$, $I_2$, and $I_3$ are defined in $(-t_w, t_w]$, $(t_w, 1+t_w)$, and $[1-t_w, 1+t_w)$ ,respectively.
They are 0 outside.
Note that the second interval overlaps with the third one.
The constant $\beta$ can be a complex number depending on the piece.
For example, the first piece of the L--F model, $\beta$ is a complex number, and $\Im[p_h(t)]$ provides the antialiased result.
For $k=0$, the zeroth-order antialiased polynomial pulse is applicable.

We conducted a set of tests using this revised L-F model.
Similar to Fig.~\ref{aaFLspecSlice880Nuttall11},
the noise level around the fundamental component was about $-180$~dB from the peak harmonics level.
This revised model is also available as a set of open access MATLAB functions.


\section{Conclusions}

We formulated and implemented a procedure to generate aliasing-free glottal source model output.
We antialiased the Fujisaki--Ljungqvist model using a newly designed cosine series antialiasing function,
followed by an IIR digital equalizer.
We also revised our antialiased L--F model using the six-term cosine series and the IIR equalizer.
The proposed procedure is general enough to be applicable to other glottal source models and
any signal models consisting of polynomial and complex exponential segments.
These antialiased models are available as open access MATALB procedures with
interactive GUI tools for education and research in speech science\cite{kawahara2016isST}.
The antialiased glottal excitation signals also provide a reliable and flexible means to test $f_{\rm o}$ extractors and
source aperiodicity analysis procedures.

\section{Acknowledgments}

This work was supported by JSPS KAKENHI Grant Numbers~JP15H03207, JP15H02726 and JP16K12464.

\bibliographystyle{IEEEtran}
\bibliography{IS2017kawaharaAAFjL}

\appendix

\section{Supplement}
The main text was accepted for publication in Proc. of Interspeech2017.
This appendix provides materials that were dropped owing to paucity of space.

\subsection{Test signals}\label{ss:testsignal}
We used the frequency-modulated $f_{\rm o}$ trajectory $f_{0i}(t)$ for generating the figures shown in this article.
The following equation provides the details:
\begin{align}
f_{0i}(t) & = f_{base} 2^{\left(\frac{f_d}{1200} \sin(2 \pi f_m t)\right)} ,
\end{align}
where $f_d$ determines the depth of the vibrato in terms of the musical cent,
and $f_m$ determines the rate of the vibrato.
We used $f_d=6$ (cent) and $f_m= 5.2$ (Hz) to generate the figures. 
The Fujisaki--Ljungqvist model parameters $\{A, B, C, R, F, D\}$ were set as follows:
{\small
\begin{align}
\!\!\!\{\!A, B, C, R, F, D\!\} \! = \!\{0.2, -1, -0.6, 0.48, 0.15, 0.12\} .
\end{align}}

We also generated test signals using the antialiased L--F model.
The same frequency-modulated $f_{\rm o}$ trajectory was used.
The L--F model parameters $\{ t_p, t_e, t_a, t_c\}$ were set as follows:
\begin{align}
\{ t_p, t_e, t_a, t_c\} & = \{ 0.4134, 0.5530, 0.0041, 0.5817\} ,
\end{align}
where the specific values are the average value of the modal voices reported in Reference\cite{childers1995jasa}.

\subsection{Spectrogram of Fujisaki--Ljungqvist model}
\begin{figure}[tbp]
\begin{center}
\ifthenelse{\value{IsDraftmode} < 1}{%
\includegraphics[width=0.99\hsize]{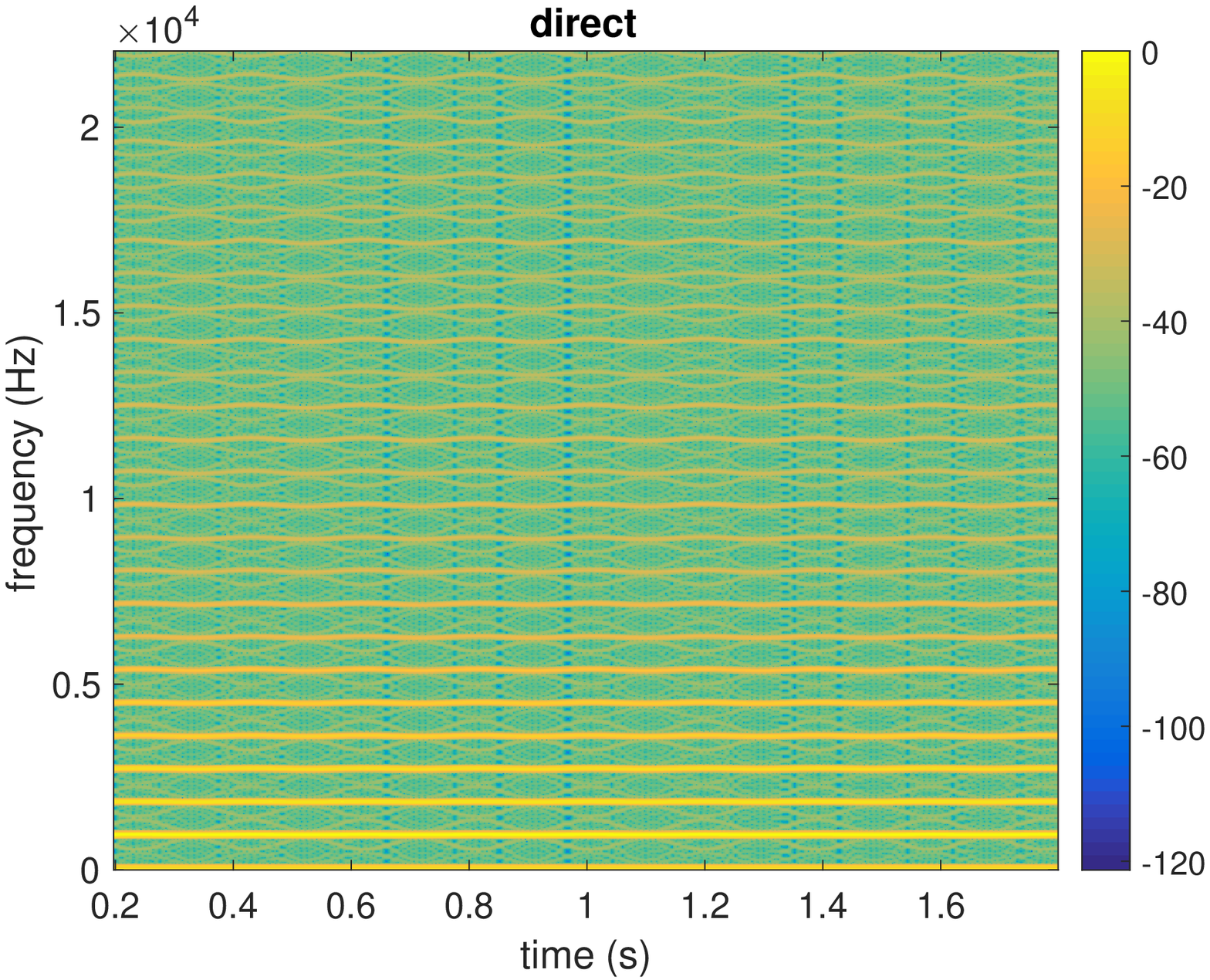}
\includegraphics[width=0.99\hsize]{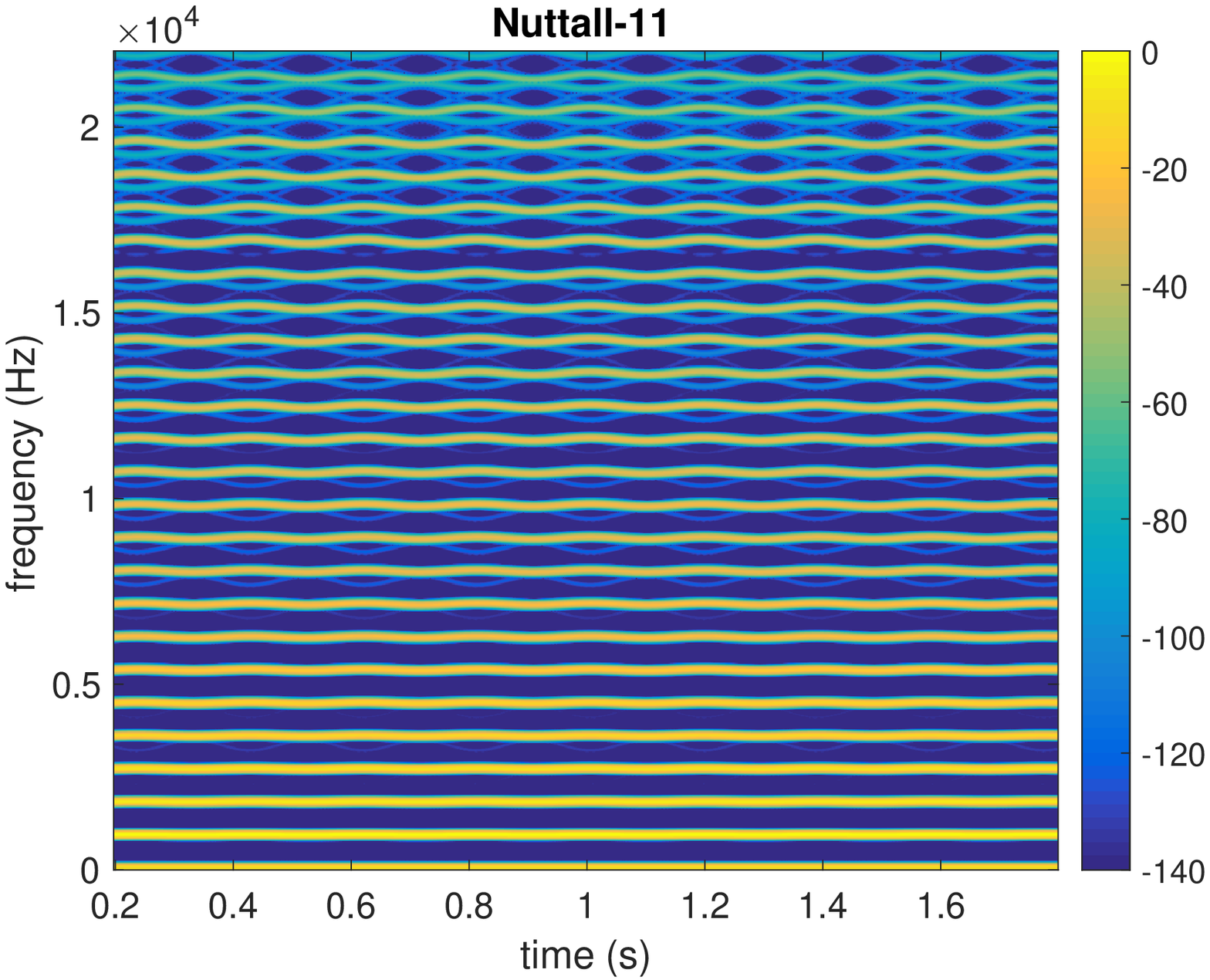}
\includegraphics[width=0.99\hsize]{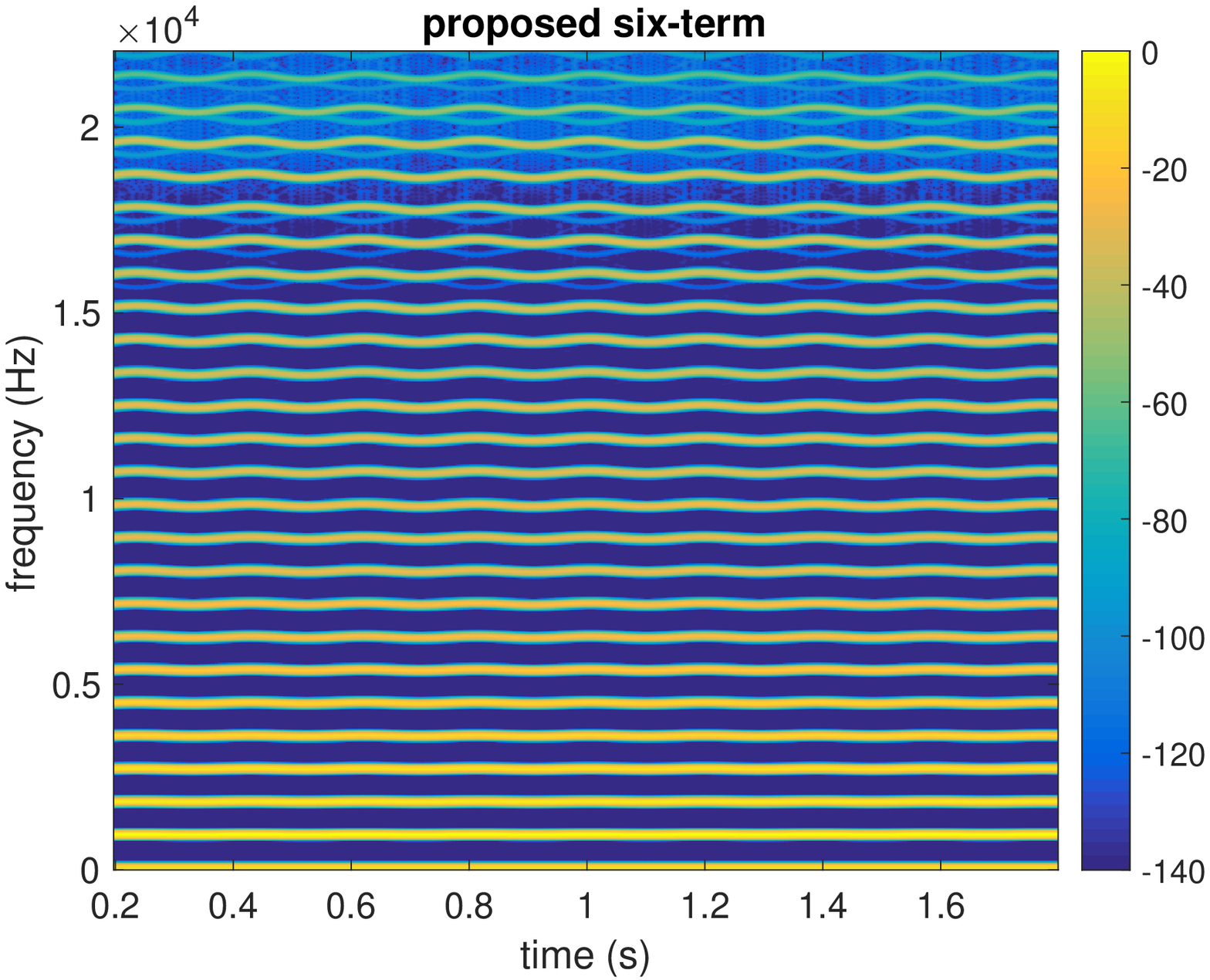}
}{%
\includegraphics[width=0.99\hsize]{figures/aaFLmodelDirectSgramIS.eps}
\includegraphics[width=0.99\hsize]{figures/aaFLmodelOldSgramIS.eps}
\includegraphics[width=0.99\hsize]{figures/aaFLmodelRevSgramIS.eps}
}
\caption{
Spectrogram of direct discretization, antialiasing with Nuttall-11 window,
and antialiasing with the proposed six-term cosine series.
}
\label{aaFLmodelRevSgramIS}
\end{center}
\end{figure}
Figure~\ref{aaFLmodelRevSgramIS} shows the spectrogram of the
generated excitation source signals using direct discretization, the Nuttall-11 window,
and the proposed six-term cosine series.
The spectrum slices in Fig.~\ref{aaFLspecSlice880Nuttall11} are
sampled from these spectrograms.

We used a self-convolved version of a Nuttall's window for calculating these spectrograms.
It is the 12th item of Table II of Reference\cite{nattall1981ieee}.
This self-convolved window has the maximum sidelobe level at $-186.64$~dB and
a 36~dB/oct decay rate.
The low sidelobe level and the steep decay rate of this window allow us to inspect low-level spurious of the proposed procedure.
The window length and frame shift were
40~ms and 2~ms, respectively.

\subsection{IIR equalizer design}
The target equalizer shape is designed to equalize attenuation up to 68~dB for
six-term and 58~dB for five-term cosine series, respectively.
The FFT buffer length was 32,768, and the
length of the FIR response was 161 taps.
The original equalizer shape, represented as an absolute spectrum, was
converted to the FIR response and truncated using
one of the Nuttall windows\cite{nattall1981ieee}.
The 12th item of Table~II of the reference was used.
It has a maximum sidelobe level of $-93.32$~dB and a decay rate of 18~dB/oct.

The IIR equalizer was designed using the autocorrelation
coefficients of this truncated FIR response, by applying LPC analysis.
\begin{figure}[tbp]
\begin{center}
\includegraphics[width=0.99\hsize]{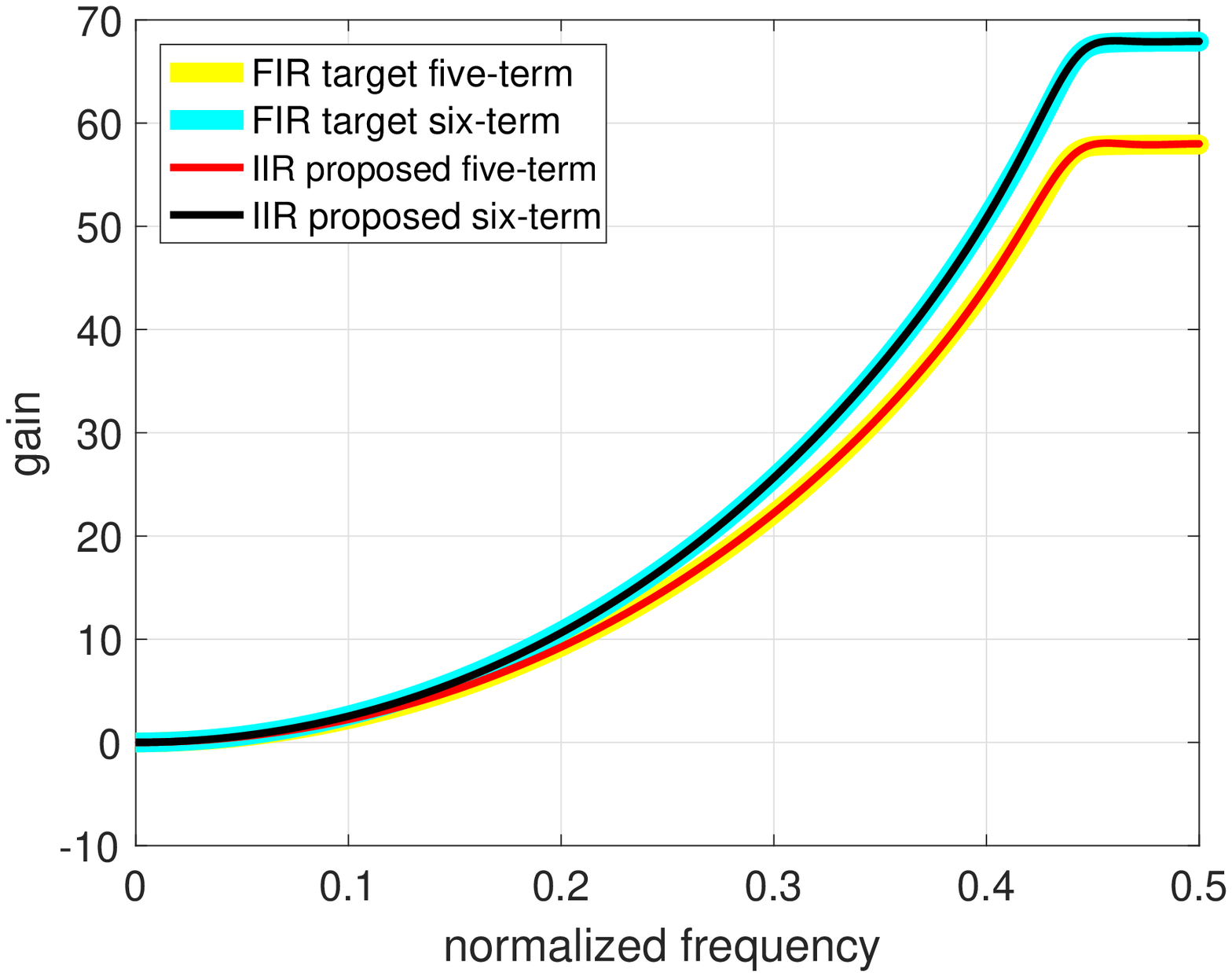}
\includegraphics[width=0.99\hsize]{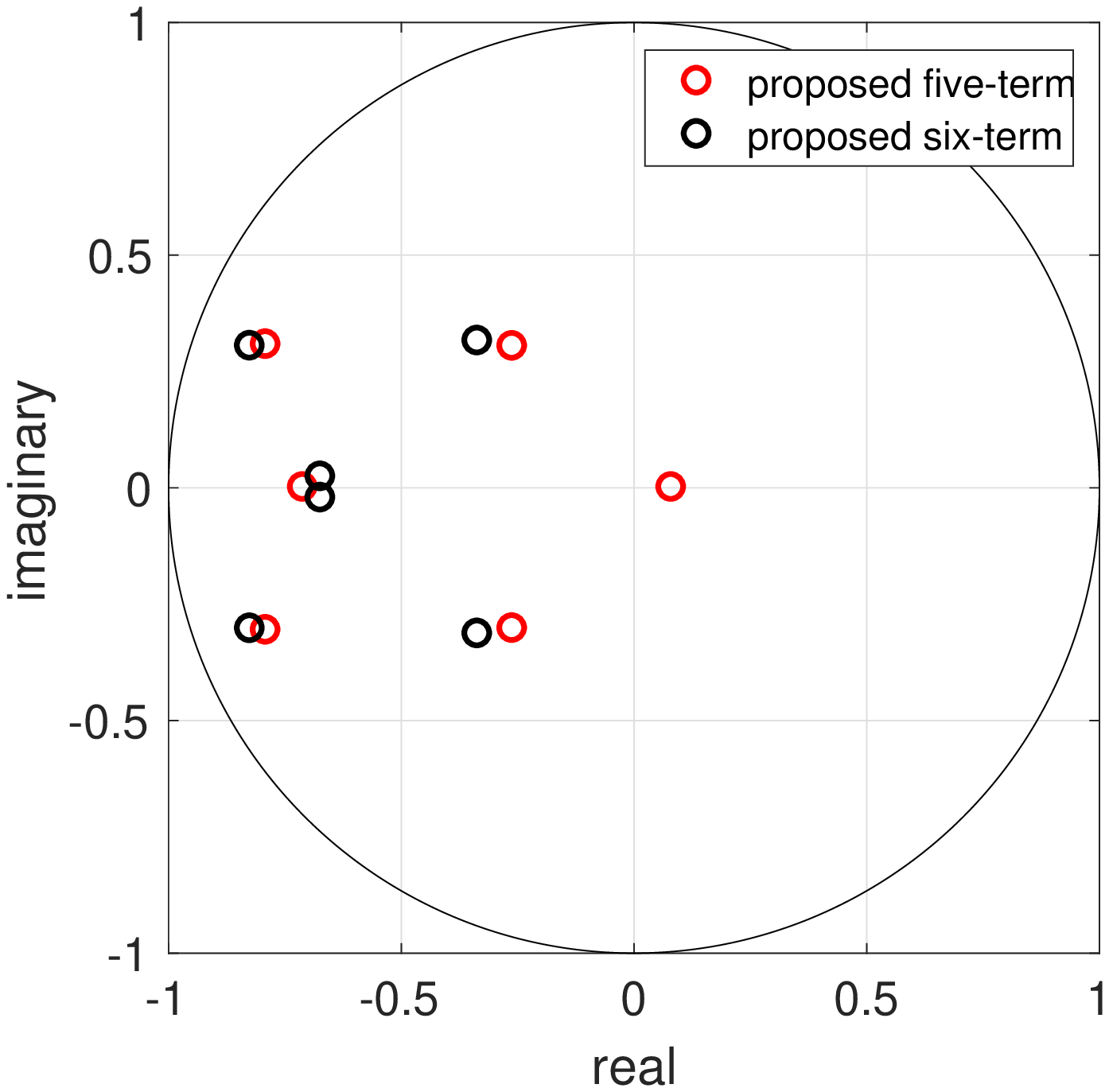}
\caption{(Upper plot) 
Frequency gain of equalizer using FIR implementation and IIR implementation.
(Bottom plot) Pole locations of each equalizer. It also shows the unit circle on the complex plane.
}
\label{equalizerImp}
\end{center}
\end{figure}
Figure~\ref{equalizerImp} shows the frequency response of each equalizer and the pole locations.
The poles are not close to the unit circle, indicating the numerical stability of the equalizers.

\subsection{MATLAB functions}
The following MATLAB functions are prepared to generate the excitation source signals.
\begin{description}
\item[\texttt{antiAliasedPolynomialSegmentR} ] Generate a time normalized segment of an antialiased polynomial pulse.
\item[\texttt{antialiasedFLmodelSingleR}] Generate one cycle of an antialiased F-L model excitation signal.
\item[\texttt{AAFjLjmodelFrom$f_{\rm o}$TrajectoryR}] Generate an antialiased F-L model excitation signal using
the given $f_{\rm o}$ trajectory and constant F-L model parameters.
\item[\texttt{AAFjLjmodelFrom$f_{\rm o}$TrajectoryTVR}] Generate an antialiased F-L model excitation signal using
the given $f_{\rm o}$ trajectory and time varying F-L model parameters.
\end{description}
We used \texttt{AAFjLjmodelFrom$f_{\rm o}$TrajectoryR} to generate the test signal used to draw Figs.~\ref{aaFLspecSlice880Nuttall11} and~\ref{aaFLmodelRevSgramIS}.

\begin{figure}[tbp]
\begin{center}
\includegraphics[width=0.99\hsize]{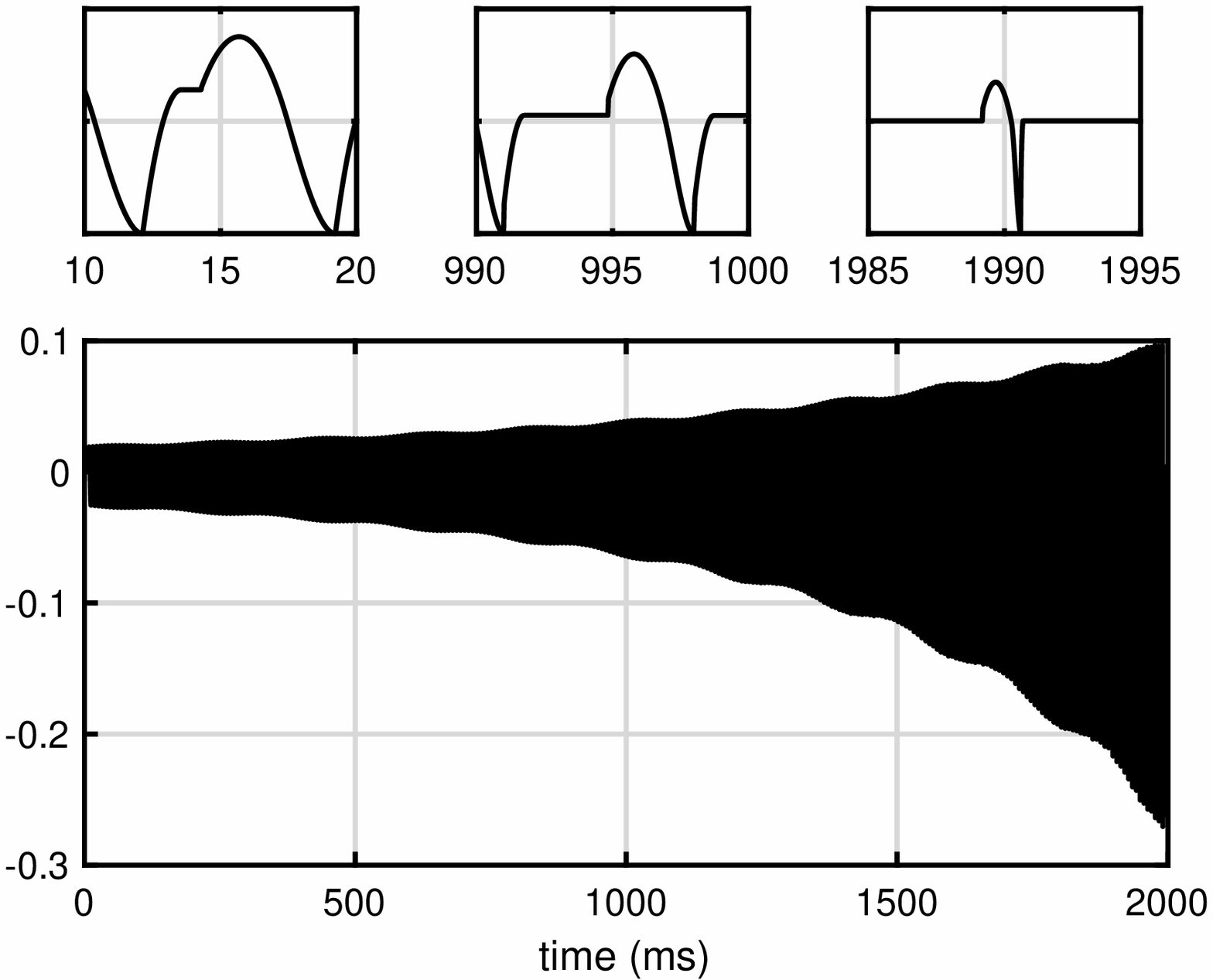}
\caption{Excitation signal generated with time varying F-L model parameters.
}
\label{morphedSourceFL}
\end{center}
\end{figure}
Figure~\ref{morphedSourceFL} shows an example of an excitation signal with time varying Fujisaki--Ljungqvist model parameters.
The total airflow of each pitch cycle is kept constant.

\subsection{Application to aliasing-free L--F model}
\begin{figure}[tbp]
\begin{center}
\includegraphics[width=0.99\hsize]{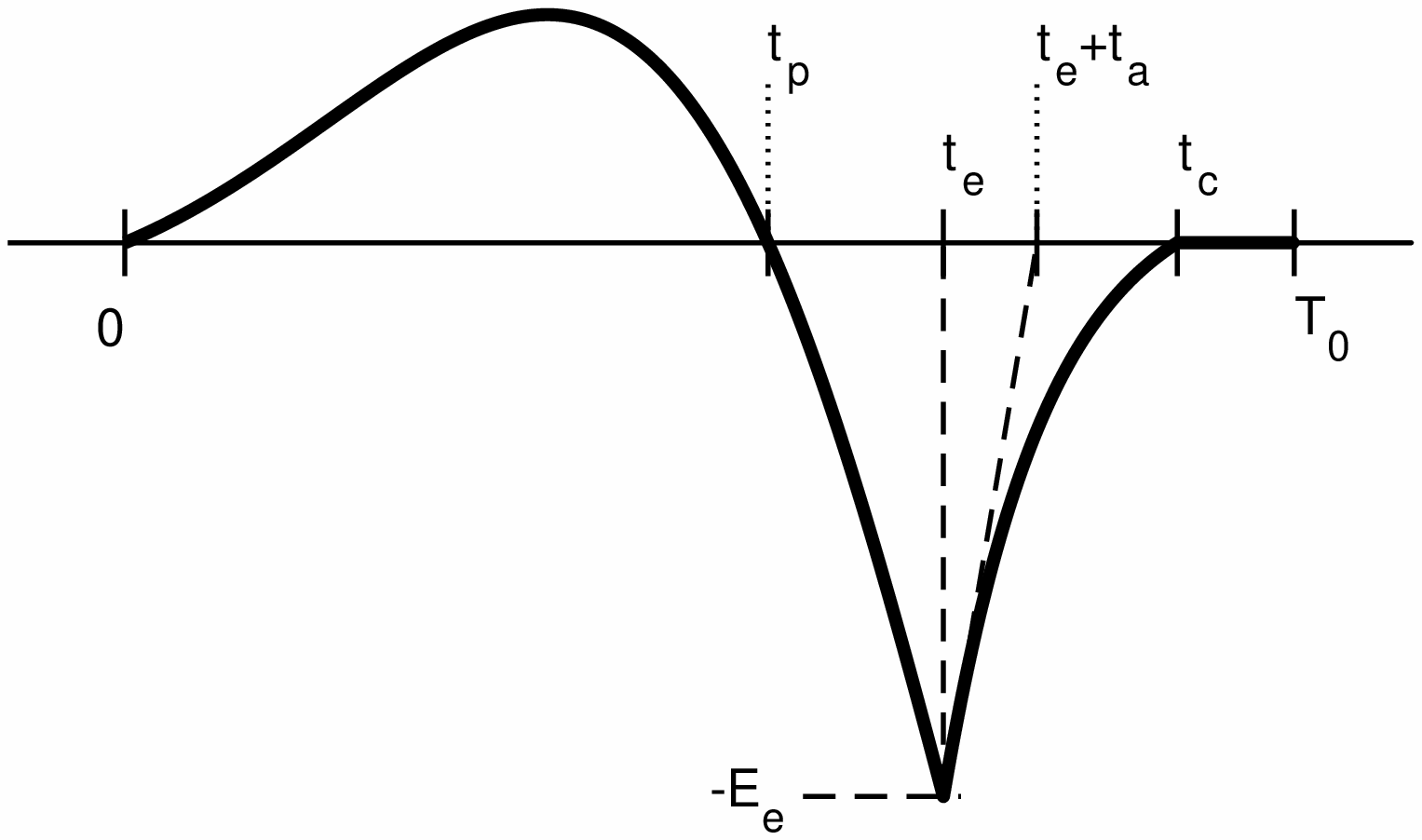}
\caption{L--F model parameters.
}
\label{lfModelParametersDef}
\end{center}
\end{figure}
Figure~\ref{lfModelParametersDef} shows the L--F model waveform with
time parameters to define it.
There is a typo in the L--F model definition given on page 6 of the original reference\cite{fant1985four}, while other descriptions in the reference are correct.
The fixed equation that defines the L-F model is as follows:
\begin{align}
E(t) & = E_0 e^{\alpha t}\sin \omega_g t &  (t<t_e) \label{eq:opening} \\
E(t) & = \frac{-E_e}{\beta t_a}  \!\left[ e^{-\beta (t-t_e)}\! -  \!e^{-\beta (t_c-t_e)}\right] & (t_e \le t < t_c) , \label{eq:closing}
\end{align}
where $E(t)$ is defined as the time derivative of the glottal airflow $U_g(t)$.
It is convenient to normalize the time axis by $T_0$ and the amplitude by $E_e$ without loss of generality.
The coefficients that can then be determined from the design parameters are $E_0/E_e, \alpha, \omega_g$, and $\beta$.
Because the airflow is zero while the vocal fold is closed, the following constraint holds:
\begin{align}
\int_0^{T_0} E(t) dt = 0 . \label{eq:zeroflow}
\end{align}

The following steps provide the parameter values.
First,  substitute $t = t_c$ in Eq.~\ref{eq:closing}.
Solving it yields $\beta$.
Then, use Eq.~\ref{eq:zeroflow} and $\omega_g = t_p/\pi$ to determine $\alpha$.
We used the numerical optimization function \texttt{fzero} of MATLAB to implement these steps.

\subsubsection{Spectrum slice and spectrogram}
\begin{figure}[tbp]
\begin{center}
\includegraphics[width=0.99\hsize]{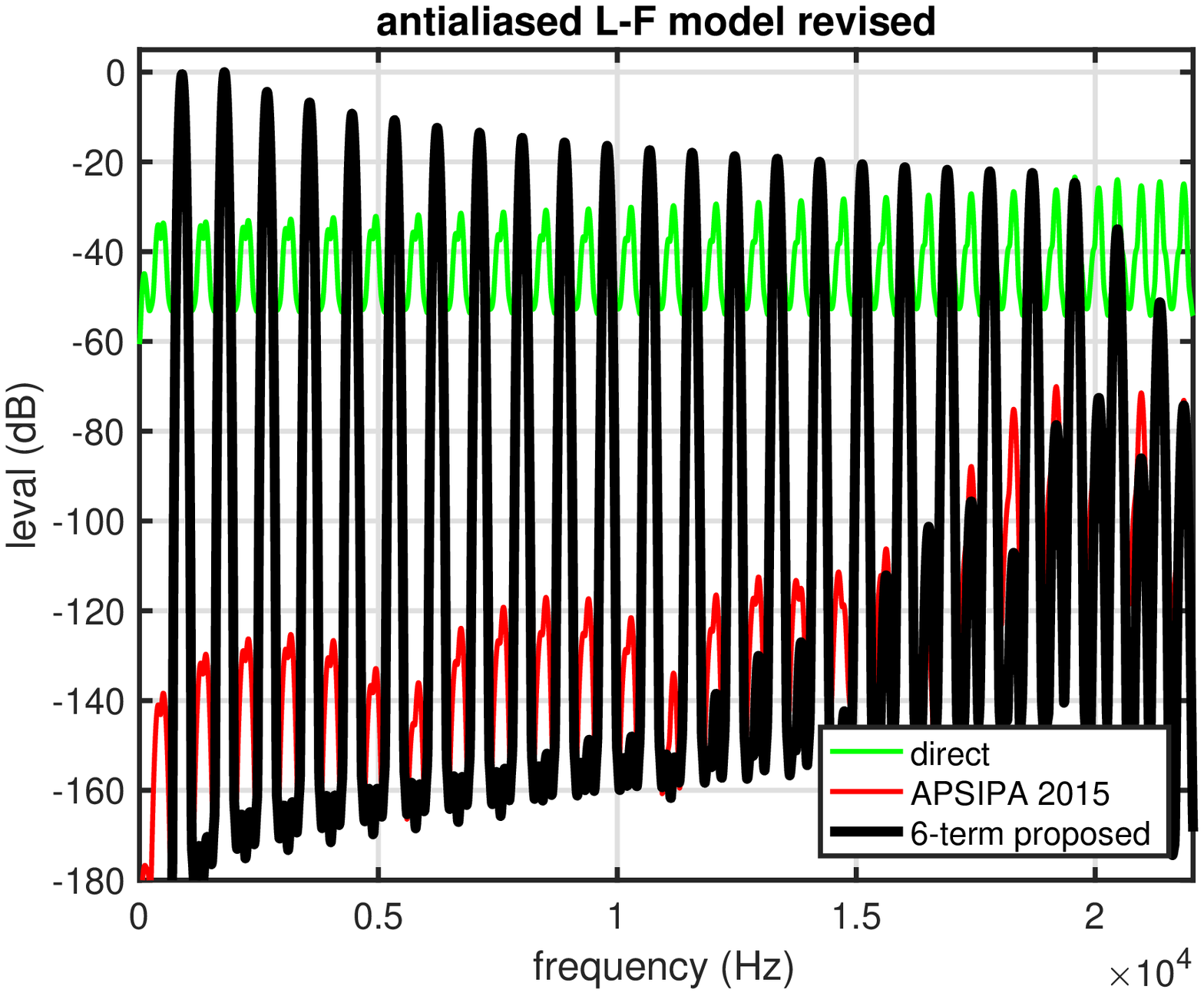}
\includegraphics[width=0.99\hsize]{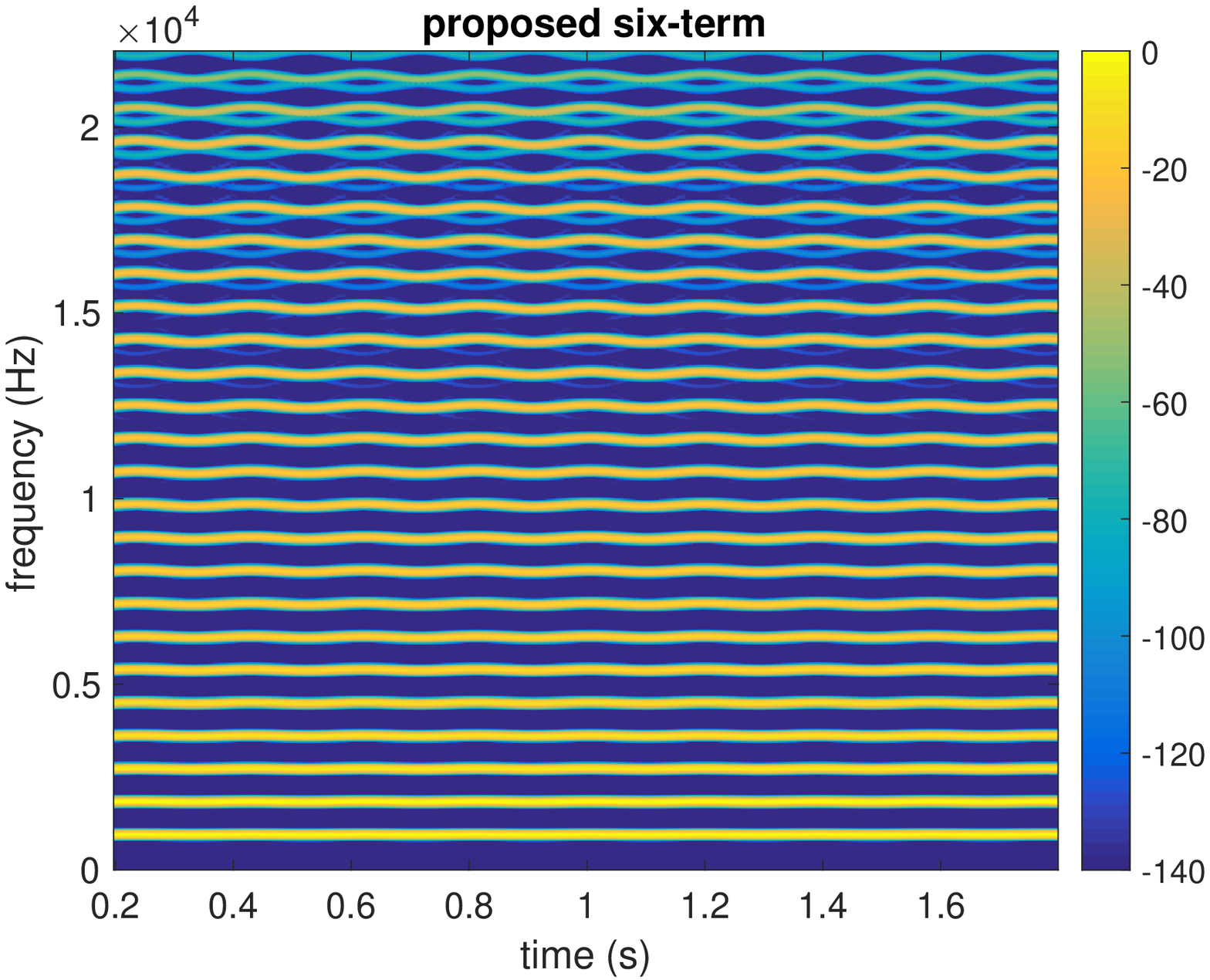}
\caption{(Upper)
Spectrum slice of direct discretization, our previous implementation,
and antialiasing with the proposed six-term cosine series.
(Lower) Spectrogram.
}
\label{aaLFmodelRevSgramIS}
\end{center}
\end{figure}
Figure~\ref{aaLFmodelRevSgramIS} shows spectrum slice and spectrogram of
a generated test signal.
The same frequency-modulated $f_{\rm o}$ trajectory was used
and the details are given in Appendix~\ref{ss:testsignal}

\end{document}